# Temporal and Conceptual Modeling: Foundations and Research Evolution

Nina Edelweiss and José Palazzo M. de Oliveira[0000-0002-9166-8801]

Federal University of Rio Grande do Sul (UFRGS), Brazil
Postgraduate Program in Computing (PPGC)
ninaedelweiss@gmail.com, palazzo@inf.ufrgs.br

**Abstract.** This paper examines the development and subsequent influence of temporal and conceptual modeling research conducted by Nina Edelweiss and José Palazzo Moreira de Oliveira. It traces a sequence of studies from 1993 to 1997 that extended object-oriented modeling with temporal data types, historical states, bitemporal queries, multi-model integration, roles, and schema evolution. The analysis then considers graduate research supervised at UFRGS, showing how these foundations supported work on temporal query processing, indexing, version management, schema versioning, adaptive systems, information integration, and evolving knowledge. By connecting the original TF-ORM framework with later research trajectories, the paper identifies a coherent progression from formal temporal database models toward broader approaches to dynamic and knowledge-based information systems. These contributions anticipated current concerns in provenance, traceability, ontology evolution, knowledge graphs, and semantically consistent adaptive systems.

## 1 Introduction

Information systems rarely remain static. Objects change state, organizational rules evolve, schemas are revised, and knowledge is continuously expanded or reinterpreted. Conventional database models, however, were originally designed primarily to represent the current state of information. This limitation motivated research on temporal databases, which seek to preserve and query the histories of data, and on conceptual models capable of expressing how objects, roles, relationships, and schemas evolve over time. Within this context, object-oriented modeling offered a particularly suitable foundation because it combined structural representation with identity, behavior, and abstraction mechanisms.

The research developed by Nina Edelweiss, José Palazzo Moreira de Oliveira, and their collaborators constitutes a coherent contribution to this field. The initial studies introduced temporal constructs into an object-oriented model [1], formalized the retrieval of valid-time and transaction-time information [2], integrated temporal object-oriented modeling with established methods for industrial applications [3], and extended the resulting framework to conceptual schema evolution through the notion of roles [4]. Together, these publications established a progression from temporal representation and querying to model integration and structural evolution.

This theoretical foundation was subsequently developed through graduate research at the Federal University of Rio Grande do Sul. Work supervised by Nina Edelweiss addressed temporal query implementation [5], state-transition rules in TF-ORM [6], temporal indexing [7], object versioning [8], and temporal schema versioning [9, 10]. In parallel, research supervised by José Palazzo Moreira de Oliveira examined evolving information in multi-agent resource allocation [11], textual information extraction [12], Web-based scientific evaluation [13], and academic collaboration networks [14]. Although the latter studies did not focus directly on temporal database management, they treated evolution, adaptation, and changing information as fundamental properties of information systems.

This paper presents a historical and conceptual synthesis of these related research trajectories. Its objective is to identify how the initial temporal object-oriented framework evolved across publications and supervised studies, as well as how its central concerns anticipated later developments in versioned data, provenance, semantic integration, ontology evolution, adaptive systems, and knowledge graphs. Rather than treating the works as isolated contributions, the analysis emphasizes their methodological continuity and their shared concern with preserving consistency and meaning as data and knowledge change.

The remainder of the paper is organized as follows. Section 2 introduces the temporal extension of the object-oriented paradigm proposed at CAiSE '93. Section 3 examines the temporal query formalization presented at DEXA '94. Section 4 discusses the integration of DFD, E-R, and object-oriented methods in the BASYS '96 industrial application. Section 5 addresses schema evolution through the role concept developed for CLEI '97. Section 6 reviews temporal database research supervised by Nina Edelweiss, while Section 7 examines research related to temporal information supervised by José Palazzo Moreira de Oliveira. Finally, Section 8 synthesizes the contributions and discusses their broader impact.

## 2 Temporal Extension of the Object-Oriented Paradigm (CAiSE '93)

The foundation of this research program was established in the paper *An Object-Oriented Temporal Model* presented at CAiSE '93 (Edelweiss, Palazzo & Pernici, 1993). The study identifies a critical limitation of conventional database modeling: the inability to represent the dynamic evolution of objects and their states. Traditional models capture only the current configuration of data, losing the historical dimension that is essential to understanding system behavior. To address this gap, the authors proposed a temporal extension of the F-ORM (Functional-ity in Object with Roles Model), creating what they called the Temporal F-ORM model.

The proposal introduced four modeling constructs: (i) a comprehensive set of temporal data types and associated functions; (ii) timestamps linked to both object instances and dynamic properties; (iii) a special null value for attributes outside their validity period; and (iv) temporal conditions expressed through temporal logic. By treating events as the primitive temporal elements, the model could represent discrete or continuous time, transaction time, and valid time. This extension enabled the formal representation of the object's life cycle and behavioral evolution an essential advance toward temporally aware information systems.

## 3 Temporal Query Formalization (DEXA '94)

The subsequent work, *An Object-Oriented Approach to a Temporal Query Language*, presented at DEXA '94 (Edelweiss, Palazzo & Pernici, 1994), focused on the retrieval dimension of temporal information. Building upon the Temporal F-ORM, the authors defined a temporal query language for object-oriented databases capable of handling bitemporal information data characterized simultaneously by transaction time and valid time. The paper proposed a taxonomy of temporal queries that classifies queries according to the combination of selection and projection components (data, time, and mixed).

This classification was systematically mapped to the different database histories, snapshot, valid-time, transaction-time, and bitemporal, allowing precise formal reasoning about what kinds of information could be retrieved from each. The proposed language retained the general structure of SQL while incorporating temporal operators inspired by TQuel. Through this development, the authors provided a framework that unifies temporal representation and retrieval, bridging the conceptual model with operational query semantics.

## 4 Multi-Modeling and Integration in Industrial Systems (BASYS '96)

In *Multi-Modeling of an Industrial Application with DFD, E-R and Object-Oriented Methods*, presented at BASYS '96 (Edelweiss & Palazzo, 1996), the theoretical framework was extended to the domain of industrial information systems. The paper addressed the challenges of modeling complex industrial environments, where different perspectives functional, structural, and behavioral must coexist. The authors proposed a multi-modeling approach combining three methods: Data Flow Diagrams (DFD) for dynamic behavior, Entity-Relationship (E-R) models for static data structures, and the TF-ORM (Temporal Functionality in Objects with Roles Model) for formal and temporal specification.

In this architecture, the TF-ORM serves as the formal substrate integrating the previous modeling layers. It unifies the system's behavioral constraints and temporal properties, enabling translation to extended relational systems such as POSTGRES or to object-oriented databases. The study demonstrates that temporal object-oriented models are not confined to academic abstraction but can effectively underpin computer-aided engineering and production control applications, ensuring consistency across conceptual, design, and operational levels.

## 5 Schema Evolution and the Concept of Role (CLEI '97)

The culmination of this line of research appears in *Evolução de Esquemas Conceituais: o Conceito de Papel*, presented at CLEI '97 (Edelweiss, Palazzo & Kunde, 1997). This work extends the TF-ORM framework to encompass conceptual schema evolution, a crucial issue in systems that must adapt to changing organizational or legal requirements. The authors argue that instead of redefining entire classes, evolution should be managed through the role concept, which encapsulates the various behaviors an object may assume during its lifetime.

In the proposed model, an object remains an instance of a single class but can dynamically instantiate, suspend, or destroy roles that represent different behavioral perspectives. This approach allows for consistent adaptation of both schema and stored data while preserving historical continuity. The paper also introduces a taxonomy of schema temporalities instantaneous, transaction-time, valid-time, and bitemporal schemas showing that the TF-ORM naturally accommodates these variants. By linking temporal database theory with schema versioning, the authors provided an elegant solution to the long-standing problem of maintaining coherence between schema evolution and data history.

## 6 Research on Temporal Databases Supervised by Nina Edelweiss

An analysis of the master's dissertations and doctoral theses supervised by Professor Nina Edelweiss, reveals the establishment of an important research line on temporal databases developed between the late 1990s and the early 2000s. Although the individual works address different aspects of the problem, they exhibit a clear scientific continuity, with each study extending previous results and collectively forming a coherent research program on the representation, manipulation, and evolution of temporal data.

One of the earliest contributions in this research line was developed by Carvalho [5], who investigated the implementation of query processing for an object-oriented temporal data model. The study focused on mechanisms capable of retrieving historical object states by explicitly incorporating the temporal dimension into the data model. Its objective was to enable queries that considered not only the current values of objects but also their evolution over time, a fundamental characteristic of temporal database systems.

Subsequently, Hübler [6] expanded the representation of temporal information through the implementation of state transition rules for the TF-ORM model.

This work addressed the formalization of object state changes and their historical persistence, providing mechanisms to preserve the temporal evolution of information while maintaining consistency. The research consolidated conceptual aspects of temporal modeling and demonstrated their feasibility through a practical implementation.

As temporal models became increasingly sophisticated, improving query performance emerged as a significant research challenge. In this context, Lehnen [7] proposed a conceptual framework to support the development of indexing techniques specifically designed for temporal databases. Rather than introducing a single indexing structure, the author defined an architectural framework intended to guide the construction of different access mechanisms while considering the unique characteristics of temporal queries and the continuous evolution of data.

Another important development in this research line concerned the integration of temporal support with version management. Gelatti [8] proposed an extension to the ODMG standard for object-oriented databases, incorporating explicit support for both temporal dimensions and persistent object versioning. This proposal anticipated issues related to the coexistence of multiple versions of the same information while simultaneously preserving its temporal history.

The evolution of database schemas themselves constituted another research direction pursued by the group. Santos [9] conducted a comparative study of schema versioning mechanisms in temporal databases, analyzing different strategies for supporting structural changes without compromising the integrity of historical information. Complementing this work, Jantsch [10] presented the implementation of the TVMSE model, providing a computational realization of temporal schema versioning and supporting the controlled evolution of database structures.

Taken together, these studies reveal a remarkably coherent scientific progression. The initial research concentrated on temporal data representation and query processing; subsequent work addressed performance issues through temporal indexing; and later studies focused on the evolution of both objects and database schemas by incorporating version management mechanisms. This sequence demonstrates an increasing concern with preserving the history of information at multiple levels of abstraction, ranging from stored data values to the database schema itself.

From a historical perspective, this research line proved particularly innovative. Many of the problems investigated during this period have regained prominence in contemporary information systems research, especially in areas such as data provenance, traceability, auditing, versioned databases, knowledge graphs, and data governance. Consequently, the work supervised by Nina Edelweiss represents a significant contribution to the development of Brazilian research on temporal databases, anticipating challenges that have since become central to distributed systems, data science, and knowledge-based information systems.

## 7 Research Related to Temporal Information Supervised by José Palazzo Moreira de Oliveira

The master's dissertations and doctoral theses supervised by Professor Palazzo reveals a research trajectory that differs significantly from the temporal database line developed by Nina Edelweiss. Rather than focusing on temporal database management systems themselves, Palazzo's research has addressed temporal aspects as one dimension of broader problems involving conceptual modeling, information integration, semantic representation, knowledge evolution, and adaptive information systems.

The earliest work directly connected to temporal information is the doctoral dissertation by Bastos [11], which investigated multi-agent systems for dynamic resource allocation. Although the primary contribution lies in distributed decision making, the proposed market-oriented mechanisms depend on the continuous evolution of system states and on adaptive allocation strategies over time, introducing temporal reasoning as an intrinsic component of autonomous coordination.

Another important contribution is the dissertation by Zambenedetti [12], which proposed techniques for information extraction from textual databases. The work focused on integrating heterogeneous information sources and organizing knowledge extracted from evolving collections of documents. While not explicitly addressing temporal databases, the research dealt with information whose value changes continuously as new textual sources become available, emphasizing the dynamic evolution of knowledge repositories.

More recently, Palazzo's supervision expanded toward Web-based information systems and scientific knowledge management. Jouris [13] investigated methods for evaluating scientific conferences through Web visibility indicators. Since these indicators naturally evolve over time, the proposed evaluation framework implicitly incorporates temporal dynamics into the assessment of scientific quality, reflecting the changing impact of academic events.

Similarly, Lopes [14] proposed methods for evaluating and recommending collaborations in academic social networks. The research considered scientific collaboration as an evolving phenomenon in which researchers, publications, and relationships continuously change over time. Recommendation strategies therefore rely not only on the current structure of collaboration networks but also on their temporal evolution, making time an essential element of the proposed models.

Viewed collectively, these studies reveal that temporal aspects in Palazzo's research emerge primarily through the concepts of evolution, adaptation, and knowledge dynamics rather than through explicit temporal data models. This perspective became increasingly aligned with subsequent research on conceptual modeling, ontologies, Semantic Web technologies, and knowledge graphs, where temporal evolution constitutes an inherent characteristic of knowledge representation rather than a separate database management problem. Consequently, the notion of time is treated as an integral property of evolving information sys-tems, supporting adaptive behavior, semantic consistency, and knowledge maintenance.

From a historical perspective, this research trajectory anticipated several themes that have become increasingly important in contemporary Artificial Intelligence and Knowledge Engineering. Current challenges involving knowledge graph evolution, provenance, continuous knowledge integration, adaptive recommender systems, and ontology evolution all rely on mechanisms capable of representing information that changes over time. Although these works did not explicitly investigate temporal database systems, they contributed to establishing conceptual and computational foundations for representing the dynamic nature of information in intelligent information systems.

## 8 Synthesis and Impact

Taken together, the studies examined in this paper reveal a sustained progression from the formal representation of temporal behavior to the broader problem of managing evolving information and knowledge. The foundational publications developed between 1993 and 1997 established the main conceptual trajectory: the CAiSE '93 paper introduced temporal behavior into an object-oriented model; DEXA '94 extended this foundation with formal query mechanisms; BASYS '96 demonstrated its applicability in integrated industrial modeling; and CLEI '97 incorporated conceptual schema evolution through the role construct. This sequence connected temporal data types, object histories, query semantics, multi-model integration, and schema evolution within a coherent methodological framework.

The dissertations supervised by Nina Edelweiss subsequently transformed these foundations into a broader research program on temporal database technology. They addressed historical query processing, the implementation of TF-ORM state transitions, temporal indexing, the integration of time and object versioning, and the controlled evolution of database schemas. Collectively, these works extended temporal modeling across multiple levels of abstraction, from stored object states and access mechanisms to persistent versions and schema histories. They also demonstrated the practical feasibility of concepts initially introduced in the earlier theoretical models.

The research supervised by José Palazzo Moreira de Oliveira followed a complementary trajectory. Instead of concentrating on temporal database management, it treated time through the evolution, adaptation, and integration of information in multi-agent systems, textual repositories, Web-based evaluation, and academic social networks. This broader perspective positioned temporal change as an inherent property of intelligent and knowledge-based information systems. The two supervisory trajectories are therefore connected by a shared concern with preserving meaning and consistency as data, structures, relationships, and knowledge evolve.

Viewed as a whole, this body of work anticipated contemporary research challenges involving data provenance, traceability, auditing, versioned databases, se-mantic interoperability, ontology evolution, adaptive recommender systems, and knowledge graph maintenance. By combining temporal logic, object orientation, conceptual modeling, version management, and knowledge evolution, Edelweiss and Palazzo helped establish foundations that remain relevant to the design of dynamic, adaptive, and semantically consistent information systems.

**Acknowledgments.** This study was partially funded by the Brazilian Conselho Nacional de Desenvolvimento Científico e Tecnológico (CNPq) and by the Coordenação de Aperfeiçoamento de Pessoal de Nível Superior – Brasil (CAPES) – Finance Code 001.

**Disclosure of Interests.** The authors have no competing interests to declare that are relevant to the content of this article. During the drafting of this paper, LLMs were utilised to assist in editing drafts written by the authors. The authors then edited and refined the text further using Grammarly. Such use and this acknowledgement adhere to the ethical guidelines for the use of generative AI in academic research. The authors have developed the work entirely, which has been thoroughly vetted for accuracy, and assumes responsibility for the integrity of their contributions.